\documentclass[a4paper,twoside]{article}
\usepackage{amsmath,amssymb,amsthm}

\newcommand{\iid}{{\mathrm{Id}}}

\newcommand{\Tc}{T_c}

\newcommand{\tc}{t_c}

\newcommand{\Rcal}{{\cal R}}

\newcommand{\Ocal}{{\cal O}}

\newcommand{\counit}{{\varepsilon}}

\newcommand{\Deltap}{\Delta'}

\newcommand{\Deltapu}{{\underline{\Delta}'}}

\newcommand{\ix}[1]{{}_{\scriptscriptstyle(#1)}}

\newcommand{\ipu}[1]{{}_{\scriptscriptstyle(\underline{#1}')}}
\newcommand{\iipu}[1]{{}_{\scriptstyle(\underline{#1}')}}

\begin{document}
% 
%\begin{titlepage}
% 
\title{\textbf{Coalgebras and quantization}}
\author{Christian Brouder\\
Institut de Min\'eralogie et de Physique des Milieux Condens\'es, 
CNRS UMR7590,\\
Universit\'es Paris 6 et 7, IPGP, 4 place Jussieu,\\
F-75252 Paris Cedex 05, France.
}%
\maketitle
\begin{abstract}
Two coalgebra structures are used in quantum field theory.
The first one is the coalgebra part of a Hopf algebra
leading to quantization. The second one is
a co-module co-algebra over the first Hopf algebra
and it is used to define connected chronological
products and renormalization.
Paper written for the Encyclopaedia of Mathematics.
\end{abstract}

%\end{titlepage}

\underline{Co-algebra} is a pervasive structure of quantum field theory. 
It enters the quantization of fields, the definition
of the chronological product and of renormalization.
According to the \underline{deformation quantization} point of view,
quantum fields are classical functions whose product
is deformed \cite{Dito90}. This deformation
can be described by a \underline{co-quasi-triangular}
structure on the \underline{Hopf algebra} of 
\underline{normal products}
\cite{Fauser,BrouderQG,Hirshfeld}. 

\section{The Hopf algebra of normal products}
Taking the example of
a scalar field, we start from the \underline{co-algebra} $C$
generated as a vector space by the \underline{Wick powers}
$\varphi^n(x)$, where $n$ is a nonnegative integer 
($x$ is a point of $\mathbb{R}^d$). The coproduct of $C$ is
given by
\begin{eqnarray*}
\Delta_C \varphi^n(x) &=& \sum_{k=0}^n \binom{n}{k} \varphi^k(x)
  \otimes \varphi^{n-k}(x),
\end{eqnarray*}
its co-unit is $\counit_C(\varphi^n(x))=\delta_{n,0}$.

The coalgebra structure of $C$ enables us to define a commutative
and cocommutative \underline{bialgebra}
$B$ which is the \underline{symmetric algebra} $S(C)$ as an algebra,
equipped with the coproduct $\Delta$ defined by
$\Delta u=\Delta_C u$ if $u\in S^1(C)=C$ and extended to
$B$ by algebra morphism (i.e.  $\Delta (uv)=\Delta u \Delta v$ 
for any $u$ and $v$ in $B$). The co-unit $\counit$ of $B$ is
defined similarly by $\counit(u)=\counit_C(u)$
if $u\in S^1(C)=C$ and extended to
$B$ by algebra morphism (i.e.  $\counit (uv)=\counit(u) \counit(v)$ 
for any $u$ and $v$ in $B$).
This product is called the \underline{normal product} or
\underline{Wick product} in quantum field
theory. The coproduct of basis elements of $B$ is
\begin{eqnarray*}
\Delta \varphi^{n_1}(x_1)\dots\varphi^{n_p}(x_p)
&=&
\sum_{k_1=0}^{n_1}\dots\sum_{k_p=0}^{n_p} \binom{n_1}{k_1}
\dots  \binom{n_p}{k_p}
 \varphi^{k_1}(x_1)\dots\varphi^{k_p}(x_p) \otimes
 \varphi^{n_1-k_1}(x_1)\dots\varphi^{n_p-k_p}(x_p).
\end{eqnarray*}
Note that the co-unit is the vacuum expectation value \cite{BrouderQG}:
$\counit(u)=\langle 0| u | 0\rangle$.

The bialgebra $B$ is transformed into a \underline{connected}
Hopf algebra 
$H$ through a quotient by the ideal (and \underline{co-ideal})
generated by elements of the form $a-\counit_C(a)1$, where 
$a\in C$ and 1 is the unit of $S(C)$.

\section{Quantization}
Quantization is now achieved by defining a 
\underline{co-quasi-triangular structure} (co-QTS) on $H$.
In $H$, a co-QTS $\Rcal$ is entirely determined by the value
of $\Rcal(a,b)$ for $a$ and $b$ in $C$.
Two co-QTS are used:
the operator co-QTS determined by
$\Rcal(\varphi^m(x),\varphi^n(y)) =
  \delta_{m,n} n! D_+(x-y)^n,$
where $D_+(x)$ is the free Wightman function and
the chronological co-QTS determined by
$\Rcal(\varphi^m(x),\varphi^n(y)) =
  \delta_{m,n} n! D(x-y)^n,$
where $D(x)$ is a regularized version of 
the Feynman propagator,
or Green function of the free field
(see \underline{quantum field theory}).

A co-quasi-triangular structure generates a
twisted product \cite{Majid} defined by
$u \circ v = \sum \Rcal(u\ix1,v\ix1) u\ix2 v\ix2.$
This expression is called Wick's theorem in
quantum field theory. When the co-QTS is
determined by the Wightman function, the twisted product
is equivalent to the star product of deformation quantization
\cite{Hirshfeld}. When the co-QTS is determined by
the Feynman propagator, the twisted product is
commutative. It is called
the \underline{chronological} or time-ordered product
or $T$-product and it plays a basic role in the
\underline{perturbation theory} of quantum fields.
More generally, for $a_1,\dots,a_p$ in $C$, the
chronological product is
$T(a_1 \dots a_p)=a_1\circ\dots\circ a_p$.
For any $u\in H$, $T(u)=\sum t(u\ix1)u\ix2$,
where $t(u)=\counit(T(u))$ is given by
\begin{eqnarray}
t(\phi^{n_1}(x_1)\circ\dots \circ \phi^{n_p}(x_p))
&=& n_1!\dots n_p!
\sum_{M} \prod_{i=1}^{p-1}\prod_{j=i+1}^p
   \frac{D(x_i,x_j)^{m_{ij}}}{m_{ij}!},
\label{t(u)}
\end{eqnarray}
where the sum is over all symmetric $p\times p$ matrices $M$ of nonnegative
integers $m_{ij}$ such that $\sum_{j=1}^p m_{ij}=n_j$ and
$m_{ii}=0$ for all $i$. Each matrix $M$ is the 
\underline{adjacency matrix} of a graph which is called
a \underline{Feynman graph}.

\section{A second co-algebraic structure}
There is a second co-product on $H$ defined by
$\Deltap u= u\otimes 1 + 1\otimes u$ if $u\in C$,
extended to $H$ by algebra morphism.
For example
\begin{eqnarray*}
\Deltap \varphi^{n_1}(x_1)\varphi^{n_2}(x_2)
&=&
\varphi^{n_1}(x_1)\varphi^{n_2}(x_2)\otimes 1+
1\otimes\varphi^{n_1}(x_1)\varphi^{n_2}(x_2)+
\varphi^{n_1}(x_1)\otimes \varphi^{n_2}(x_2)
\\&&+ \varphi^{n_2}(x_2)\otimes\varphi^{n_1}(x_1).
\end{eqnarray*}
More generally
$\Deltap \varphi^{n_1}(x_1)\dots\varphi^{n_p}(x_p)$
is determined by taking all subsets 
$I$ of $\{1,\dots,p\}$
and defining
\begin{eqnarray*}
\Deltap \varphi^{n_1}(x_1)\dots\varphi^{n_p}(x_p)
&=&
\sum_I \Big(\prod_{i\in I} \varphi^{n_i}(x_i)\Big)
\otimes \Big(\prod_{j\notin I} \varphi^{n_j}(x_j)\Big),
\end{eqnarray*}
with the convention that $\prod_{i\in I}\varphi^{n_i}(x_i)=1$ if 
$I=\varnothing$. This second coproduct is
very natural and it has been implicitly used for
a long time in physics \cite{Ruelle,Epstein}.

We denote by $(H,\Deltap)$ the co-algebra which is equal
to $H$ as a vector space, with co-product $\Deltap$
and co-unit $\counit$ (the co-unit of the Hopf algebra $H$).
The co-algebra $(H,\Deltap)$ and the Hopf algebra $H$
have an important relation: $(H,\Deltap)$
is a co-module co-algebra over $H$ \cite{Majid}. In
other words, $(H,\Deltap)$ is a 
\underline{co-module} over $H$ (with
the \underline{co-action} $\psi=\Delta$) satisfying the compatibility 
property
\begin{eqnarray*}
(\Deltap\otimes\iid)\psi &=&
(\iid\otimes\iid\otimes\mu)(\iid\otimes\tau\otimes\iid)
(\psi\otimes\psi)\Deltap,
\end{eqnarray*}
where $\tau(u\otimes v)=v\otimes u$
and $\mu(u\otimes v)=uv$.

The co-product $\Deltap$ enables us to define
the connected and the renormalized chronological products.
The \emph{reduced co-product} is 
$\Deltapu u= \Deltap u - u\otimes 1 - 1 \otimes u$,
its iteration is
$\Deltapu^{(0)}=\iid$, 
$\Deltapu^{(1)}=\Deltapu$, 
$\Deltapu^{(n+1)}=(\Deltapu\otimes\iid^{\otimes n})\Deltapu^{(n)}$, 
and its action on $u\in H$ is denoted by
\begin{eqnarray*}
\Deltapu^{(n-1)} u &=&
\sum u\ipu1\otimes\dots\otimes u\ipu{n}.
\end{eqnarray*}

The connected chronological
product \cite{Epstein} is now defined by
\begin{eqnarray*}
\Tc(u) &=& -\sum_{n=1}^\infty \frac{(-1)^n}{n} 
   T(u\ipu1) {\dots} T(u\iipu{n}),
\end{eqnarray*}
for $u\in\ker\counit$.
Because $(H,\Deltap)$ is a co-module co-algebra over $H$,
$\Tc(u)=\sum \tc(u\ix1) u\ix2$, where $\tc(u)$ is
given by eq.(\ref{t(u)}) with a sum over adjacency matrices
$M$ corresponding to \underline{connected} graphs.
The renormalized chronological product is defined by
\cite{Bogoliubov,PinterHopf}
\begin{eqnarray*}
T_R(u) &=& \sum_{n=1}^\infty \frac{1}{n!}
  T\big(\Ocal(u\ipu1)\dots \Ocal(u\iipu{n})\big),
\label{TpdeT}
\end{eqnarray*}
for $u\in\ker\counit$, where
$\Ocal$ is a linear map from $H$ to $C$
called a \emph{generalized vertex} \cite{Bogoliubov}.
The renormalized chronological product implements the
\underline{renormalization} of quantum field theory.

Note that a similar construction is possible in a noncommutative
context where the symmetric algebra $S(C)$ is replaced by
the tensor algebra $T(C)$. The second co-product $\Deltap$
is then the deconcatenation co-product.

\end{document}